\documentclass[twocolumn, aps, amsfonts, amsmath, showpacs]{revtex4}

\usepackage{graphicx}

\begin{document}

\title{Correlation effects in electronic structure of PuCoGa$_5$}

\author{L. V. Pourovskii$^1$, M. I. Katsnelson$^1$, and A. I. Lichtenstein$^2$}
\affiliation{$^1$Institute for Molecules and Materials, Radboud University of Nijmegen,
NL-6525 ED Nijmegen, The Netherlands \\
$^2$Institut f\"ur Theoretische Physik, Universit\"at Hamburg,
20355 Hamburg, Germany}

\date{\today}

\begin{abstract}

We report on results of the first realistic electronic structure calculations of 
the Pu-based PuCoGa$_5$ superconductor
based on the dynamical mean field theory. We find that
dynamical correlations due to the local Coulomb interaction
between Pu $f$-electrons lead to substantial modification of the
electronic structure with a narrow peak being formed in vicinity
of $E_F$, in agreement with the experimental photoemission
spectra, and in contrast with the recent calculations within 
the LDA+U method \cite{Shick2005},
where only static electronic correlations have been included. 
Both Pu and Co contribute in equal footing to the narrow peak on the density
of states at the Fermi level, the Co partial density of states
being prominently affected by electronic correlations on the Pu
sites.  The ${\bf k}$-resolved spectral density is calculated and 
the theoretical spectral function resolved extended Van Hove singularity near $E_F$. This singularity may lead to enchancement of the magnetic susceptebility
and favour $d-$wave superconductivity.

\end{abstract}

\maketitle

The recent discovery of superconductivity in PuCoGa$_5$
\cite{Sarr2002} and PuRhGa$_5$ \cite{Wast2003} has attracted a lot
of attention to electronic structure of these Pu-based compounds.
They have the same HoCoGa$_5$-type tetragonal structure as the
Ce-based superconductors Ce$M$Ga$_5$ ($M$ is transition metal)
\cite{Mathur98}, however PuCoGa$_5$ has an order of magnitude
higher transition temperature $T_c=$18.5 K, the highest $T_c$
among known $f$-electron superconducting materials. Both
Ce$M$Ga$_5$ and Pu$M$Ga$_5$ are unconventional superconductors
\cite{Curro2005} with the nodal $d-$ wave symmetry of
superconducting gap $\Delta$ and antiferromagnetic fluctuations
being a likely mechanism for their unconventional
superconductivity. It has been suggested \cite{Bauer2004} that the
increase of $T_c$ in the Pu$M$Ga$_5$ compounds in comparison with
Ce$M$Ga$_5$ might be due to more itinerant character of the 5$f$ band
in Pu.

The experimental data \cite{Sarr2002} show no evidence of a
long-range magnetic structure down to temperature of 1 K, whereas
the magnetic susceptibility obeys the modified Curie-Weiss law
with an effective local moment 0.68 $\mu_B$. The specific heat
coefficient $\gamma$=77 mJ mol$^{-1}$K$^{-2}$ \cite{Sarr2002} is
slightly larger than the value observed in $\delta$-Pu (50 mJ
mol$^{-1}$K$^{-2}$ \cite{Wick80}) indicating possible enhancement
by strong dynamical spin fluctuations. The experimental
photoemission spectra (PES) of  PuCoGa$_5$ \cite{Joyce2003}
exhibits a broad feature centered 1.2 eV below the Fermi level,
which is made up of Co $d$- and Pu $f$-bands, as well as a narrow
peak at the Fermi level. The latter feature is similar to narrow
resonances observed in the PES of $\delta$-Pu \cite{Havel2003} as
well as in the spectra of some Pu-based compounds
\cite{Goud2000,Dur2004}.

The electronic structure of PuCoGa$_5$ has been investigated in
several papers on the basis of the local spin density
approximation (LSDA)
\cite{Opahle2003,Maeh2003,Soder2004,Opahle2004}. In the
non-spin-polarized band structure calculations of Refs.
\onlinecite{Opahle2003,Maeh2003} for nonmagnetic state the main
$f$-state spectral weight is located in the vicinity of the Fermi
level in contradiction with experiment. The spin polarization pushes 
occupied part of the $f$-band below the Fermi level
\cite{Opahle2003,Soder2004}, with the density of states (DOS) for
the antiferromagnetic phase \cite{Soder2004} being in good
agreement with the experimental PES \cite{Joyce2003}. However, the
Pu total magnetic moment obtained within the LSDA is too large
(around 2 $\mu_B$) in comparison with experimentally measured
$\mu_{eff}$ as well as the ordered magnetic structure in
PuCoGa$_5$  predicted by the LSDA calculations has not been
observed in experiment. Joyce {\it et al.} \cite{Joyce2003}
obtained theoretical PES in a very good agreement with the
experimental one using the mixed level model (MLM). In the MLM
5$f$ electrons of Pu is divided in four localized and one
itinerant, however the energy position of the localized states is
not calculated  {\it ab initio} but rather treated as a parameter
which value is chosen to match experiment.

The strong enhancement of the specific heat coefficient $\gamma$,
as well as the characteristic narrow resonance at the Fermi level,
hint on possible importance of dynamical electronic correlations
in PuCoGa$_5$. In $\delta$-Pu \cite{kotliarNature} and Pu
monochalcogenides \cite{Pour2005} a good agreement between the
theoretical and measured PES for non-magnetic phase (which is the
ground state for these materials in accordance with the
experiment) has been obtained only when dynamical correlations
between $f$-electrons were taken into account by means of the
dynamical mean-field theory (DMFT) \cite{Geor96} in a framework of
the LDA+DMFT \cite{AnisDMFT,lda++} method. For PuCoGa$_5$ Shick
{\it et al.} \cite{Shick2005} have taken into account only {\it
static} correlations by means of the LDA+U method. They obtained
the antiferromagnetic ground state with the reduced (in comparison
with the LSDA) Pu effective magnetic moment around 1 $\mu_B$. At
the same time, the density of states reported in Ref.
\onlinecite{Shick2005} demonstrates some discrepancy with the
experimental PES. While the position of the broad manifold
reproduced quite well, the narrow feature at $E_F$ is absent. It
seems reasonable to suggest that by analogy with $\delta$-Pu and
other Pu-based compounds an adequate description of the electronic
structure of PuCoGa$_5$ requires dynamical correlations to be
taken into account. It is important to notice that in the Pu-based compounds
the dynamical correlations affect
most profoundly the states in vicinity of the Fermi level \cite{Pour2005}, 
which are involved in superconductivity. It is the purpose of this Letter to present
electronic structure of PuCoGa$_5$ with the dynamical electronic
correlations explicitly included within the LDA+DMFT approach.

We start with the LSDA+U calculations of PuCoGa$_5$. Following
Ref. \onlinecite{Shick2005} to allow for the antiferromagnetic
ground state we double the unit cell along the $\hat a$ and $\hat
b$ crystallographic axes and point magnetic moments of the
nearest-neighbor Pu atoms in the Pu planes in opposite directions,
the magnetic moments of Pu atoms being aligned along the
crystallographic $\hat c$ axis. The calculations have been carried
out by the full-potential linear MT-orbitals method (FPLMTO)
\cite{Savr96} at experimental lattice parameter of PuCoGa$_5$
($a$=7.845 \AA, $c/a$=1.603 for the unit cell) with the spin-orbit
coupling included in a self-consistent second-order variational
procedure. We include the full matrix of on-site Coulomb interaction between Pu
5$f$-electrons parametrized  by the average Coulomb interactions $U=$3 eV and the
Hunds-rule exchange $J=$=0.55 eV. These values of $U$ and $J$ are in the
range of commonly accepted values for Pu. For the double counting
(DC) term we have employed so-called ``around mean-field''
formulation of the LDA+U method \cite{Anis91,Czyz94}. This version
of the LDA+U method was recently successfully used to explain
non-magnetic ground state of the Pu $\delta$-phase \cite{ShickEL}.
We used previously  the same values of $U$ and $J$ as well as the
same choice of the DC in the DMFT simulations of Pu
monochalcogenides \cite{Pour2005}.

\begin{figure}
\includegraphics[width=0.4\textwidth, angle=270]{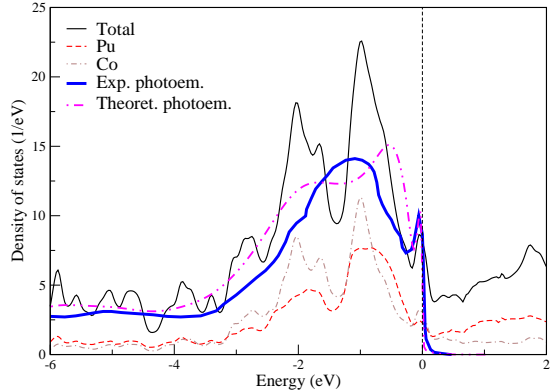}
\caption{The total DOS (solid line) as well as the partial DOS for
Pu (dashed line) and Co (dot-dashed line) atoms obtained within
the DMFT. The thick solid and dot-dashed lines are the
experimental \cite{Joyce2003} and theoretical PES in arbitrary
units, respectively. } \label{SPTF_dos}
\end{figure}

The one-particle Hamiltonian $H_t({\bf k})$ obtained by the
converged LDA+U calculations was then orthogonalized by the
L\"owdin transformation \cite{Lowd50} and used to compute the
local Green's function by means of the Brillouin zone (BZ)
integration
\begin{equation}\label{G_loc}
G(i\omega)=\sum_{\bf k} [(i\omega+\mu){\bf 1}-H_t({\bf
k})-\Sigma^{dc}(i\omega)]^{-1},
\end{equation}
where $\omega=(2n+1)\pi T$ are the fermionic Matsubara frequencies
for a given temperature $T$, $\mu$ is the chemical potential and
$\Sigma^{dc}(i\omega)$ is the local self-energy with a ``double
counting'' term subtracted. Within the DMFT scheme the local
self-energy has been obtained by the solution of the many-body
problem for a single quantum impurity  coupled to an effective
electronic ``bath'' through the Weiss field function \cite{Geor96}
\begin{equation}
{\mathcal G}^{-1}(i\omega)=G^{-1}(i\omega)+\Sigma^{dc}(i\omega).
\end{equation}

\begin{figure*}
\includegraphics[width=0.9\textwidth]{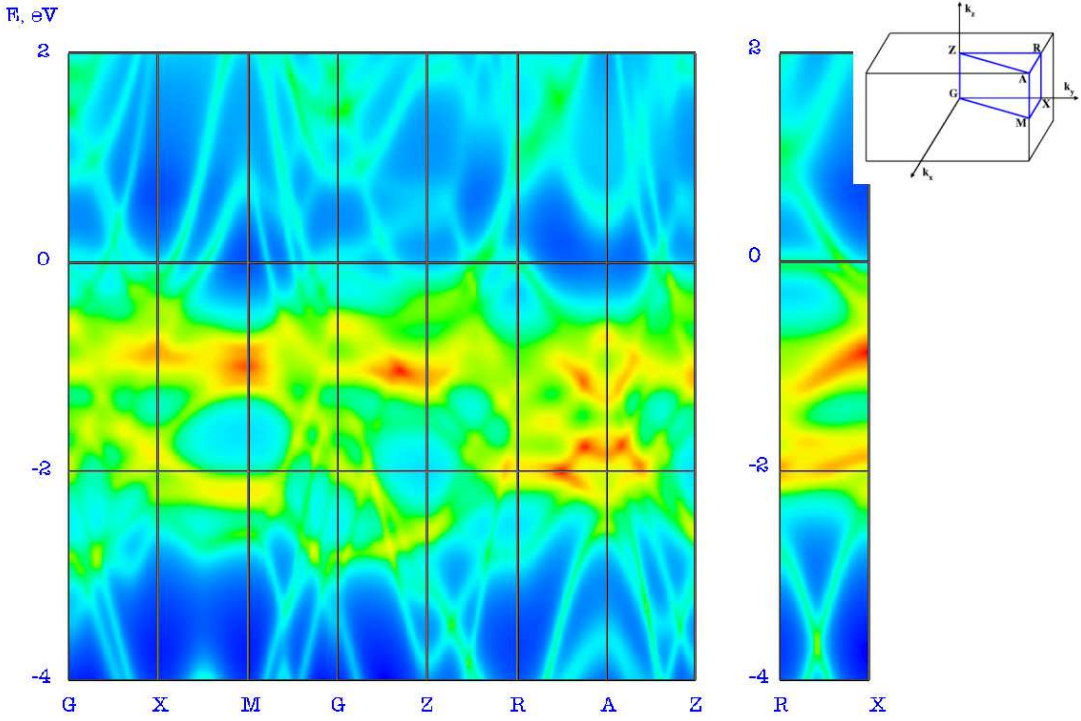}
\caption{The spectral function of PuCoGa$_5$. The sequence of colors in order of
increasing spectral density is blue, green, yellow, and red. In the inset the BZ
of PuCoGa$_5$ is shown together with our designations for the high-symmetry points.
}
\label{Spec_func}
\end{figure*}

As a quantum impurity solver we have employed the recently
developed  spin-orbit version \cite{Pour2005} of the $T$-matrix
and fluctuating exchange technique (SPTF) \cite{Kats99,Kats2002}.
In the spin-orbit SPTF the local self-energy is written as a sum
of three contributions:
\begin{equation} \label{Sigma}
\Sigma=\Sigma^{(TH)}+\Sigma^{(TF)}+\Sigma^{(PH)},
\end{equation}
where $\Sigma^{(TH)}$ and $\Sigma^{(TF)}$ are the $T$-matrix
``Hartree'' and ``Fock'' contributions, respectively,
$\Sigma^{(PH)}$ is the particle-hole contribution. $\Sigma^{(TH)}$
and $\Sigma^{(TF)}$ are obtained by substitution in the
corresponding Hartree and Fock diagrams  the bare Coulomb
interaction with the frequency dependent $T$-matrix, the latter is
given by summation of the ladder diagrams in the particle-particle
channel. $\Sigma^{(PH)}$ is obtained by the RPA-type summation in
the particle-hole channel with the bare vertex being substituted
by static limit of the $T$-matrix \cite{Kats2002,Pour2005}. While
being of perturbative type the spin-orbit SPTF solver has been
shown to provide an adequate description of correlations in 3$d$
metals \cite{Kats99,Kats2002}, half-metallic ferromagets
\cite{lulu}, and some actinide systems \cite{Pour2005}.

For the double counting term in DMFT simulations we have used the
static limit of the self energy \cite{Kats2002}. For the
spin-orbit SPTF quantum impurity solver we employed 10$^3$
Matsubara frequencies and temperature 500~K. It is important to
notice that with our choice of the double counting term the
Hartree-Fock contribution is already included into the LDA+U
one-particle Hamiltonian, therefore making magnetization almost
temperature independent. We have carried out DMTF iterations until
convergence in both the chemical potential $\mu$ and the local
self-energy was achieved. Then the Pade approximant is used
\cite{Vidb77} for analytical continuation of the local self-energy
to the real axis in order to obtain the DOS $n(E)=-1/\pi {\rm Tr}
\Im [G(E)]$ and spectral function $A(E,{\bf k})=-1/\pi {\rm Tr}
\Im[G(E,{\bf k})]$.

We started from the antiferromagnetic state of PuCoGa$_5$ but the
self-consistent calculations converged to practically zero
magnetic moments both on the Pu and Co sites. The long-range
magnetic order is suppressed due to combined influence of a strong
relativistic spin-orbit coupling and the local Coulomb interaction
effects. The $f_{5/2}$ states of Pu in PuCoGa$_5$ are almost
completely filled (the occupation number for $f$-band $n_f$=5.5)
whereas the empty $f_{7/2}$ states are located around 3 eV above
the Fermi level. It is this splitting that suppress the magnetism
of Pu in comparison with the LSDA electronic structure. The Co
3$d$ band is located between -2 eV and the Fermi energy.

The DMFT results for PuCoGa$_5$ DOS is shown on the
Fig.~\ref{SPTF_dos} together with the experimental PES spectrum
\cite{Joyce2003} and the theoretical PES obtained from the DMFT
DOS. In order to compute the theoretical PES spectrum the
calculated DOS data have been multiplied by the Fermi function
corresponding to the experimental temperature of 77 K
\cite{Joyce2003} and then convoluted with a Gaussian to simulate
for the instrumental resolution and lifetime broadening of the
core hole \cite{Joyce2003,Arko2000}. One may see a pronounced quasiparticle peak
at the Fermi level in the DMFT DOS. The theoretical PES agrees
quite well with the experimental one, the shape of the feature at
the Fermi level as well as relative spectral weight of the broad
manifold and the resonance at  $E_F$ being reproduced very
accurately. However, the manifold is somewhat wider in the
theoretical PES compare with the experiment, with a peak at -0.6
eV and a shoulder at -1.8 eV.

Interestingly, while in our DMFT calculations the correlation
effects have been taken into account only on the Pu sites, both
the Pu $f$- and Co $d$-bands are substantially modified in
comparison with the LDA+U result and they both contribute in equal
footing to the narrow peak at $E_F$. The LSDA \cite{Opahle2004} as
well as LSDA+U \cite{Shick2005} calculations for the
antiferromagnetic phase predict that the Pu $f$-band gives by far
the dominant contribution to DOS at $E_F$. We have performed also
the LDA+DMFT calculations with the local Coulomb interaction
between Co $d$-electrons being included as well (with $U$=2 eV and
$J$=0.9 eV), however that leads to rather minor changes in the
electronic structure.

The spectral function of PuCoGa$_5$ calculated within the DMFT is
displayed in Fig.~\ref{Spec_func}. The maximums in spectral
function around $M$, $A$ and $Z$ points correspond to both the Pu
$f$-band and Co $d$-band, contributing mainly to the manifold at
energies between -2 and -1 eV. The increase of the spectral weight
near the the Fermi level due to van Hove singularities (saddle
points) close to $R$ and $\Gamma$ points is mainly due to the Co
$d$-band, and in between $Z$ and $R$ points is due to
contributions from both Co and Pu. The flat Co $d$ band located
exactly at the Fermi level is clearly seen in the spectral
function along the $R-X$ direction. The extended van Hove singularity near
$R$ point together with the van Hove singularity at $\Gamma$ point result
 in strong ${\bf q}$-dependence of magnetic
susceptibility (cf., e.g., the ``parquet''
results for the two-dimensional Hubbard model \cite{IKK}). This enhanced 
susceptibility can lead to $d-$wave superconductivity \cite{BSW89}.  
Relevance of the van Hove singularities for the superconductivity
has been discussed for high-temperature cuprates \cite{IKK,mark},
as well as for Sr$_2$RuO$_4$ \cite{ruth}. However, in comparison
with these cases, the electron spectrum of PuCoGa$_5$ is a bit
less anisotropic and probably can not be considered as a purely
two-dimensional one even in a zeroth-order approximation (see the
results for $R-X$ direction in Fig.~\ref{Spec_func}).

In conclusion, we have shown that the dynamical correlation
effects modify electronic structure of the PuCoGa$_5$
superconductor in essential way. We have taken the dynamical
correlations into account by means of the dynamical mean-field
theory using the spin-orbit $T$-matrix and fluctuating exchange
(SPTF) approximation for the quantum impurity solver. in agreement
with the experimental data, the self-consistent solution turned
out to be non-magnetic. The DOS obtained by the DMFT simulations
is in a good agreement with the experimental PES, with the narrow
peak at the Fermi level being very well reproduced. While only Pu
sites have been treated as correlated, the dynamical correlations
modify the Co bands as well. On the other hand, taking into
account local correlations on Co sites does not effect noticeably
on the electronic structure. Both the Co and Pu bands contribute
almost on equal footing to the quasiparticle resonance at $E_F$,
in contrast with the LSDA picture, where only Pu bands contribute
at DOS on the Fermi level. Theoretical spectral function
demonstrates the extended Van Hove singularity near $R$ point which is due
to the Co $d$ band. New ARPES experiments are highly desirable to
investigate the energy spectrum of PuCoGa$_5$ in the vicinity of
the Fermi level and check the theoretical predictions.

\newpage

\end{document}